\documentclass[12pt,intlimits,reqno]{article}

\usepackage{amsmath}

\usepackage{enumerate}

\usepackage{amssymb}

\usepackage{amsthm}

\usepackage{bbm}
\usepackage{fancyhdr}
\usepackage{mathrsfs}
\usepackage[all]{xy}
\usepackage{graphicx}
\usepackage[polish,english]{babel}
\usepackage{verbatim}
\usepackage[utf8]{inputenc}

\newtheorem{thm}{Theorem}[section]

\newtheorem{prop}[thm]{Proposition}

\theoremstyle{definition}

\theoremstyle{remark}

\numberwithin{equation}{section}

\newcommand{\be}{\begin{equation}}
\newcommand{\ee}{\end{equation}}
\newcommand{\ol}{\overline}

\DeclareMathOperator{\Tr}{Tr}
\DeclareMathOperator{\de}{d}

\begin{document}

\title{A family of integrable perturbed Kepler systems}
\author{Anatol Odzijewicz\footnote{aodzijew@uwb.edu.pl}, Aneta Sli\.{z}ewska\footnote{anetasl@uwb.edu.pl}, Elwira Wawreniuk\footnote{ewawreniuk@math.uwb.edu.pl}}

\maketitle

\begin{center}
Institute of Mathematics\\
University of Bia\l{}ystok\\
Cio\l{}kowskiego 1M, 15-245 Bia\l{}ystok, Poland\\

\end{center}

\tableofcontents
\begin{abstract}
In the framework of the Poisson geometry of  twistor  space we consider a family of perturbed $3$-dimensional Kepler systems. We show  that Hamilton equations of this systems are integrated by quadratures. Their solutions  for some subcases are given explicitly in terms of Jacobi elliptic functions.
\end{abstract}
\section{Introduction}
It is well known that the $3$-dimensional Kepler system could be described, and thus integrated, as a system of four harmonic oscillators which are tied  by a certain quadratic bound, e.g. see \cite{iwai1,  OS}.

Here, we also consider such systems of harmonic oscillators assuming however, that they interact in a non-linear way. The interaction is given by the  Hamiltonians presented in (\ref{H1}). These Hamiltonians depend on two integer parameters $k, l \in \mathbb{Z}$ and on two arbitrary smooth functions $H_0$ and $G_0$ of four real variables. We show, see Section \ref{sec:four}, that such Hamiltonian systems are integrable. 

Next, in Section \ref{sec:Kepler}, after reduction to the submanifold of null-twistors  and applying respective  symplectic diffeomorphisms, see Proposition \ref{prop1} and Proposition \ref{prop:32}, we show that these Hamiltonian systems are equivalent to a $3$-dimensional perturbed Kepler systems written in the "fictitious time" representation, \cite{kummer, SS}. In such a way we obtain a family of  $3$-dimensional perturbed Kepler systems which are integrated  by quadratures.

 Two massive rigid bodies, which  are not spherically symmetric and interact by the Newton potential could be considered as some physical models of these perturbed Kepler systems, see the Hamilton equations (\ref{217}) and (\ref{HPKe}). 

In Section \ref{sec:ex}  particular cases of (\ref{Hequa1}) and (\ref{HPKe}) are considered  and their solutions are  explicitly presented.

\section{Preliminaries}\label{sec:Pre}

Here we describe some Poisson structures related in a canonical way to the twistor space. Twistor space $\mathbb{T}$, introduced firstly by R. Penrose \cite{penrose}, is $\mathbb{C}^4$ equipped with the scalar product 
\begin{equation}\label{product0}
\langle v, w \rangle := v^+ \phi w 
\end{equation}
of $v,w \in \mathbb{C}^4$, defined by a hermitian matrix $\phi = \phi^+ \in Mat_{4\times 4}(\mathbb{C})$, which has signature "$++--$" and satisfies $\phi^2 = \mathbbm{1}$. The symmetry group of $\mathbb{T}$ is $U(2,2)\subset Mat_{2\times 2 }(\mathbb{C})$, i.e. $g \in U(2,2)$ iff 
\begin{equation*}
g^+ \phi g = \phi. 
\end{equation*}

The Lie algebra $\mathcal{U}(2,2)$ of $U(2,2)$ consists of $\mathfrak{X} \in Mat_{4\times 4}(\mathbb{C})$ such that 
\begin{equation*}
\mathfrak{X}^+ \phi + \phi \mathfrak{X} =0. 
\end{equation*}
The $Ad(U(2,2))$-invariant pairing 
\begin{equation*}
\mathcal{U}(2,2)\times \mathcal{U}(2,2) \ni (\mathfrak{X}, \mathcal{Y}) \mapsto \Tr(\mathfrak{X}\mathcal{Y}) \in \mathbb{R} 
\end{equation*}
allow us to identify $\mathcal{U}(2,2)$ with its dual $\mathcal{U}(2,2)^*$ and define the Lie-Poisson bracket
of the linear functions 
 $\textbf{L}_\mathcal{Y} (\cdot):= \Tr ( \mathcal{Y}\cdot)$ and $\textbf{L}_\mathcal{Z} (\cdot):= \Tr ( \mathcal{Z}\cdot)$,  as follows
\begin{equation}\label{LPbracket0}
\{ \textbf{L}_\mathcal{Y} , \textbf{L}_\mathcal{Z}\}_{_{L-P}} := \textbf{L}_{[\mathcal{Y},\mathcal{Z}]}.
\end{equation}
This Lie-Poisson bracket extends to the space $C^\infty(\mathcal{U}(2,2),\mathbb{R})$ of all smooth real functions  on $\mathcal{U}(2,2)$ and 
the linear map $\textbf{L} : \mathcal{U}(2,2) \to C^\infty (\mathcal{U}(2,2), \mathbb{R})$ is a monomorphism of the Lie algebras.  For more about the theory of Lie-Poisson spaces see for example \cite{D-Z, kirillov}.

One defines the $U(2,2)$-invariant symplectic form on the twistor space $\mathbb{T}$ as $ \de \gamma$, where 
\begin{equation}\label{gamma0}
\gamma := -i v^+ \phi \de v . 
\end{equation}
The Poisson bracket of $f,g \in C^\infty (\mathbb{T}, \mathbb{R})$ corresponding to $ \de \gamma$ is 
\begin{equation}\label{pb00}
\{f, g \} ( v , v^+) = 
\sum_{k,l=1}^4 i\phi_{kl} \left(\frac{\partial g }{\partial v_k} \frac{\partial f }{\partial \ol{v}_l} - \frac{\partial f }{\partial v_k}  \frac{\partial g }{\partial \ol{v}_l} \right).
\end{equation}

The map $\textbf{J}: \mathbb{T} \to \mathcal{U}(2,2)$ defined by 
\begin{equation}\label{momentummap0}
\textbf{J}(v,v^+) := i v v^+ \phi 
\end{equation}
satisfies $\textbf{J}(gv, (gv)^+) = g\textbf{J}(v,v^+) g^+$ for $g \in U(2,2)$ and it is a Poisson map of the symplectic space $(\mathbb{T} ,  \de \gamma)$ into the Lie-Poisson space $(\mathcal{U}(2,2)\cong \mathcal{U}(2,2)^* , \{ \cdot , \cdot \}_{L-P})$, i.e. for $F,G \in C^\infty (\mathcal{U}(2,2), \mathbb{R})$ one has 
\begin{equation}
\{F\circ \textbf{J}, G\circ \textbf{J}\} = \{F, G\}_{L-P}\circ \textbf{J}. 
\end{equation}
So,  (\ref{momentummap0}) is a momentum map, e.g. see \cite{MW,Sou}, for  the $U(2,2)$-symplectic manifold $(\mathbb{T},  \de \gamma)$.  It has the following property 
\begin{equation}\label{property0}
\textbf{J}(v, v^+)^2 = i \langle v, v\rangle \textbf{J}(v,v^+)
\end{equation}
which will be useful in subsequent. Let us also note that 
\be\label{lJ} (\textbf{L}_{\mathfrak{X}}\circ \textbf{J})(v,v^+)=iv^+\phi \mathfrak{X} v,\ee
where $\mathfrak{X}\in \mathcal{U}(2,2)$.

Further on, we will use the spinor notation for the twistor space $\mathbb{T}=(\mathbb{C}^4,\phi)$, i.e.  the twistor $v\in \mathbb{T}\cong \mathbb{C}^4$ will be written  $v=\left(\begin{array}{c}\eta \\\xi\end{array}\right)\in \mathbb{C}^2 \oplus \mathbb{C}^2$ 
as a pair of spinors $\eta,\xi\in \mathbb{C}^2$. 
As a consequence of this, we will write the elements of $End \ \mathbb{T}\cong  Mat_{4\times 4}(\mathbb{C})$  in the block form $\left(\begin{array}{cc}A & B \\
C & D
\end{array}\right)$, where $A, B, C, D\in Mat_{2\times 2}(\mathbb{C})$. As a basis  in the complex vector space $ Mat_{2\times 2}(\mathbb{C})$
we will use the Pauli  matrices
\be\label{Pauli}\sigma_0 = \left(\begin{array}{cc}
1&0\\
0&1
\end{array}\right), \quad\sigma_1 = \left(\begin{array}{cc}
0&1\\
1&0
\end{array}\right), \quad\sigma_2 = \left(\begin{array}{cc}
0&i\\
-i&0
\end{array}\right), \quad\sigma_3 = \left(\begin{array}{cc}
1&0\\
0&-1
\end{array}\right),\ee
which also form the basis of the real vector space $H(2)$ of the hermitian $2\times 2$-matrices.

The following two spinor  representations $\mathbb{T}_d=(\mathbb{C}^2 \oplus \mathbb{C}^2, \phi_d)$ and $\mathbb{T}_a=(\mathbb{C}^2 \oplus \mathbb{C}^2, \phi_a)$, where
\begin{equation}\label{da}
\phi_d := \left(\begin{array}{cc}
\sigma_0 & 0 \\
0 & -\sigma_0
\end{array}\right) \quad\mbox{ and }\quad \phi_a = i\left(\begin{array}{cc}
0 & -\sigma_0\\
\sigma_0 & 0
\end{array}\right),
\end{equation}
of the twistor space $\mathbb{T}\cong (\mathbb{C}^4, \phi)$ will be used in subsequent.

These two representations, called later the diagonal and anti-diagonal representations are related by 
\begin{equation}\label{C} \left(\begin{array}{c}
\eta\\
\xi \end{array}\right)= \mathcal{C} \left(\begin{array}{c}
\vartheta\\
\zeta \end{array}\right), \qquad \phi_a = \mathcal{C}^+ \phi_d \mathcal{C},
\end{equation}
where $\eta,\xi, \vartheta, \zeta\in \mathbb{C}^2$ and the unitary map 
$\mathcal{C} : \mathbb{C}^2 \oplus \mathbb{C}^2 \to \mathbb{C}^2 \oplus \mathbb{C}^2$ is defined as follows
\begin{equation}\label{unitarytransform0}
\mathcal{C} := \frac{1}{\sqrt{2}} \left(\begin{array}{cc}
\sigma_0 & -i\sigma_0 \\
-i\sigma_0 & \sigma_0
\end{array}\right) \in U(n),
\end{equation}

Below, as in (\ref{da}), we will mark by $"d"$ the objects taken in diagonal representation and by $"a"$ the ones taken in the anti-diagonal representation.
 So, 
 the differential one-form $\gamma$, see (\ref{gamma0}), written in the spinor coordinates  is given by 
\begin{equation}\label{gammad}
\gamma_d =
-i (\eta^+ d\eta - \xi^+ d\xi )
\end{equation}
and by 
\begin{equation}\label{gammaa}
\gamma_a = \zeta^+ d\vartheta-\vartheta^+ d\zeta  , 
\end{equation}
respectively.
The group   $U_d(2,2)$ in the diagonal representation one defines as follows: 
$g=\left(\begin{array}{cc} A&B\\ C&D\end{array}\right)\in U_d(2,2)$ iff
\be A^+A=\sigma_0+C^+C,\quad D^+D=\sigma_0+B^+B \quad {\rm and}\quad D^+C=B^+A,\ee
where $A,B,C,D\in Mat_{2\times 2}(\mathbb{C})$. 

For  $\mathfrak{X}\in \mathcal{U}_d(2,2)$ one has
\be \mathfrak{X}=\left(\begin{array}{cc} i\alpha&\beta\\ \beta^+&i\delta\end{array}\right)\ee
where $\beta\in Mat_{2\times 2}(\mathbb C)$ and $\alpha,\beta\in  H(2)$. For anti-diagonal representation we have $\tilde g=\left(\begin{array}{cc} \tilde A&\tilde B\\ \tilde C&\tilde D\end{array}\right)\in U_a(2,2)$ iff 
\be \tilde A^+\tilde C=\tilde C^+\tilde A,\quad \tilde D^+\tilde B=\tilde B^+\tilde D\quad {\rm and }\quad \tilde A^+\tilde D=\sigma_0+\tilde C^+\tilde B\ee
and $\tilde{\mathfrak{X}}\in \mathcal{U}_a(2,2)$ iff
\be \tilde{\mathfrak{X}}=\left(\begin{array}{cc} \tilde\alpha&\tilde\beta\\ \tilde\gamma &-\tilde\alpha^+\end{array}\right),\ee
where $\tilde\alpha\in Mat_{2\times 2}(\mathbb{C})$ and $\tilde\beta,\tilde\gamma\in H(2)$.

The Lie-Poisson bracket (\ref{LPbracket0}) of $F,G\in C^\infty(\mathcal{U}_d(2,2),\mathbb{R})$ expressed in the matrix coordinates $(\alpha,\delta, \beta)\in H(2)\times H(2)\times Mat _{2\times 2}(\mathbb{C})$ assumes the form 
\begin{equation}\label{eq:325} 
\{F, G\}_{_{_{d,L-P}}} (\alpha, \delta , \beta , \beta^+ ) = \end{equation}
$$=-i \Tr \Bigg(\alpha\left(\left[\frac{\partial F}{\partial \alpha }, \frac{\partial G}{\partial \alpha}\right] +i\frac{\partial F}{\partial \beta }\frac{\partial G}{\partial \beta^+ } - i\frac{\partial G}{\partial \beta }\frac{\partial F}{\partial \beta^+}\right)$$
$$+ \beta\left(\frac{\partial F}{\partial \beta^+ }\frac{\partial G}{\partial \alpha } + \frac{\partial F}{\partial \delta }\frac{\partial G}{\partial \beta^+ } - \frac{\partial G}{\partial \beta^+ }\frac{\partial F}{\partial \alpha }- \frac{\partial G}{\partial \delta }\frac{\partial F}{\partial \beta^+ }\right) $$
$$+ \beta^+ \left(\frac{\partial F}{\partial \alpha }\frac{\partial G}{\partial \beta } + \frac{\partial F}{\partial \beta }\frac{\partial G}{\partial \delta } - \frac{\partial G}{\partial \alpha }\frac{\partial F}{\partial \beta }- \frac{\partial G}{\partial \beta }\frac{\partial F}{\partial \delta}\right) $$
$$+ \delta\left(\left[\frac{\partial F}{\partial \delta }, \frac{\partial G}{\partial \delta }\right] +i \frac{\partial F}{\partial \beta^+ }\frac{\partial G}{\partial \beta }- i\frac{\partial G}{\partial \beta^+ }\frac{\partial F}{\partial \beta }\right)\Bigg).$$

The momentum map (\ref{momentummap0}) in the diagonal and anti-diagonal  representation is given by 
\begin{equation}\label{diagmommap}
\textbf{J}_d (\eta, \xi, \eta^+, \xi^+) = i \left(\begin{array}{cc}
\eta\eta^+ & -\eta \xi^+ \\
\xi\eta^+ & -\xi\xi^+ \end{array}\right) 
\end{equation}
and by
\begin{equation}\label{Ja}
\textbf{J}_a (\vartheta, \zeta, \vartheta^+, \zeta^+) = \left(\begin{array}{cc}
- \vartheta \zeta^+ & \vartheta \vartheta^+\\
-\zeta \zeta^+& \zeta \vartheta^+ \end{array}\right),
\end{equation}
respectively. They are related by 
\begin{equation}
\textbf{J}_d \circ \mathcal{C} = Ad_{\mathcal{C}} \circ \textbf{J}_a, 
\end{equation}
where $\mathcal{C}:\mathbb{T}_a\to \mathbb{T}_d$ is defined in (\ref{unitarytransform0}).

Following \cite{AO} let us consider the cotangent bundle $T^*U(2)\cong U(2)\times iH(2)$ of the unitary group $U(2)$. The canonical  one-form $\pi_d$ on $T^*U(2)$ in the matrix coordinates $(Z, i\rho)\in U(2)\times iH(2)$ assumes the form 
\be \pi_d=-iTr(\rho Z^+dZ).\ee
 The group $U_d(2,2)$ acts on $U(2)\times iH(2)$ preserving the  symplectic form $\de\pi_d$ in the following way
\be \Lambda_d(g)(Z,\rho)=((AZ+B)(CZ+D)^{-1}, (CZ+D)\rho(CZ+D)^+).\ee
The corresponding momentum map $\textbf{I}_d:U(2)\times iH(2)\to U_d(2,2)$ is given by
\be \textbf{I}_d(Z,\rho)=i\left(\begin{array}{ccc} Z\rho Z^+&\ &-Z\rho\\ (Z\rho)^+&\ &-\rho\end{array}\right)\ee
and it satisfies 
\be Ad_g\circ \textbf{I}_d=\textbf{I}_d\circ \Lambda_d(g)\ee
for $g\in U_d(2,2)$.

Let us define the action $\Lambda_a:U_a(2,2)\times (H(2)\times H(2))\to H(2)\times H(2)$ of $U_a(2,2)$ on $H(2)\times H(2)$ by

\be\label{Lambda} \Lambda_a({\tilde g})(Y,X)=((\tilde A Y+\tilde B)(\tilde C Y+\tilde D)^{-1}, (\tilde C Y+\tilde D)X(\tilde C Y+\tilde D)^+)\ee
Note that $\Lambda _a({\tilde g})$ is defined correctly for  $\det (\tilde C Y+\tilde D)\not=0$ only.

The symplectic form $d\pi_a$ on $H(2)\times H(2)$, where $\pi_a$ in the coordinates $(Y,X)\in H(2)\times H(2)$ is defined by
\be\label{pia} \pi_a:=\Tr(XdY)=d(\Tr(XY))-\Tr(YdX),\ee
is invariant with respect to the action (\ref{Lambda}) and the corresponding momentum map is given by
\be\label{Ia} \textbf{I}_a(Y,X)=\left(\begin{array}{cc} -YX& YXY\\ -X&XY\end{array}\right).\ee

This momentum map satisfies the equivariance property 
\be \textbf{I}_a\circ \Lambda_a({\tilde g})=Ad_{\tilde g}\circ \textbf{I}_a\ee
for $\tilde g\in U_a(2,2)$.

The unitary matrix  (\ref{unitarytransform0}) defines
\be\label{T*}T^*_\mathcal{C}(Y,X):= ((Y-i\sigma_0)(-iY+\sigma_0)^{-1}, \frac{1}{2}(-iY+\sigma_0)X(-iY+\sigma_0)^+)\ee
the smooth one-to-one map $T^*_\mathcal{C}:H(2)\times H(2)\to U(2)\times iH(2)\cong T^*U(2)$ which is a morphism  of the vector  bundles as well as a morphism of the symplectic  manifolds.

The first component of (\ref{T*}) is the Cayley transformation
\be\label{z} Z= (Y-i\sigma_0)(-iY+\sigma_0)^{-1},\ee
which maps $H(2)$ into $U(2)$, and the second one 
\be\label{rho} \rho= \frac{1}{2}(-iY+\sigma_0)X(-iY+\sigma_0)^+\ee
is a linear authomorphism of $H(2)$ which preserves the conditions $\det X=0$ and $\Tr X\geq 0$. One obtains $U(2)$ adding to Cayley image of $H(2)$ the elements $Z\in U(2)$ satisfying the condition $\det (iZ+\sigma_0)=0$. So, the inverse of (\ref{z}), given by 
\be Y=(Z+i\sigma_0)(iZ+\sigma_0)^{-1},\ee
is defined for $\det(iZ+\sigma_0)\not=0$ only. 

We summarize the above considerations in the following diagram

 \unitlength=5mm \begin{equation}\label{diagram}\begin{picture}(11,4.6)
    \put(-2,4){\makebox(0,0){$T^*U(2)$}}
    \put(5,4){\makebox(0,0){$\mathcal{U}_d(2,2)$}}
    \put(12,4){\makebox(0,0){$\mathbb{T}_d$}}
    \put(12,0){\makebox(0,0){$\mathbb{T}_a$}}
    \put(5,0){\makebox(0,0){$\mathcal{U}_a(2,2)$}}
    \put(-2,0){\makebox(0,0){$H(2)\times H(2)$}}

    \put(-2,1){\vector(0,1){2}}
    \put(5,1){\vector(0,1){2}}
    \put(12,1){\vector(0,1){2}}
    \put(11,4){\vector(-1,0){4}}
    \put(11,0){\vector(-1,0){4}}
 \put(0.8,0){\vector(1,0){2.5}}
 \put(0,4){\vector(1,0){3}}
    \put(1.5,4.4){\makebox(0,0){$\textbf{I}_d$}}
     \put(8.5,4.4){\makebox(0,0){$\textbf{J}_d$}}
    \put(8.5,0.5){\makebox(0,0){$\textbf{J}_a$}}
\put(1.5,0.5){\makebox(0,0){$\textbf{I}_a$}}
  \put(-1,2){\makebox(0,0){$T^*_{\mathcal{C}}$}}
    \put(6,2){\makebox(0,0){$Ad_\mathcal{C}$}}
\put(12.5,2){\makebox(0,0){$ \mathcal{C}$}}
\put(-2.21,1){\makebox(0,0){$\cup$}}
    \end{picture},\end{equation}
		where $Ad_\mathcal{C}(\tilde{\mathfrak{X}}):=\mathcal{C}\tilde{\mathfrak{X}}\mathcal{C}^+$ and $\tilde{\mathfrak{X}}\in\mathcal{U}_a(2,2)$. 
		
		It is important for the subsequent considerations to mention that  all the  arrows in the diagram (\ref{diagram}) are $U(2,2)$-equivariant Poisson maps. For details see \cite{AO}.
		
		\section{System of four non-linear oscillators}\label{sec:four}

		In this section, using the diagonal representation $\mathbb{T}_d= (\mathbb{C}^4 , \phi_d )$ of the twistor space, we consider certain Hamiltonian system of four one-dimensional harmonic oscillators which are interacting in a non-linear way.

As a Hamiltonian of this system we take
\begin{multline}\label{H1}
H= H_0 (|\eta_1|^2 , |\eta_2|^2 , |\xi_1|^2 , |\xi_2|^2 ) \\
+ G_0(|\eta_1|^2 , |\eta_2|^2 , |\xi_1|^2 , |\xi_2|^2 ) (\eta_1 ^k\eta_2^{-k}\xi_1^l \xi_2^{-l} + \eta_1 ^{-k}\eta_2^{k}\xi_1^{-l} \xi_2^{l}) , 
\end{multline}
where $H_0$, $G_0$ are arbitrary smooth functions of four real arguments and $k,l\in \mathbb{Z}$. In (\ref{H1})  for $z\in \mathbb{C}$ and $k\in \mathbb{Z}$ we assumed the convention
\begin{equation}
z^k := \left\{\begin{array}{lr}
z^k, & \mbox{ if } k\geq 0 \\
\bar{z}^{-k}, & \mbox{ if } k<0 
\end{array}\right. .
\end{equation}

The Poisson bracket (\ref{pb00}) written in the diagonal representation assumes the following form
\begin{equation}\label{pb3}
\{f,g\} = i\sum_{k=1}^2\left(\frac{\partial f}{\partial \ol\eta_k}\frac{\partial g}{\partial \eta_k}-\frac{\partial g}{\partial \ol\eta_k}\frac{\partial f}{\partial \eta_k}\right) -i \sum_{k=1}^2\left(\frac{\partial f}{\partial \ol\xi_k}\frac{\partial g}{\partial \xi_k}-\frac{\partial g}{\partial \ol\xi_k}\frac{\partial f}{\partial \xi_k}\right)
\end{equation}
where $f,g \in C^\infty (\mathbb{C}^2 \oplus \mathbb{C}^2 , \mathbb{R})$.
 The  Hamilton equations defined by (\ref{H1}) are
\begin{eqnarray}\label{Hequa1}
\frac{\de}{\de t}\eta_k=i\frac{\partial H}{\partial \ol\eta_k},\qquad&\quad &\qquad \frac{\de}{\de t}\ol\eta_k=-i\frac{\partial H}{\partial \eta_k},\nonumber\\
\ &\ &\ \\
\frac{\de}{\de t}\xi_k=-i\frac{\partial H}{\partial \ol\xi_k},\qquad&\quad &\qquad  \frac{\de}{\de t}\ol\xi_k=i\frac{\partial H}{\partial \xi_k}.\nonumber
\end{eqnarray}

In order to integrate them we will use the methods investigated in \cite{GO}, \cite{AO} and \cite{KS} . For this reason, let us note that the diagonal part of the momentum map (\ref{diagmommap}) defines the momentum map $\textbf{D}_d: \mathbb{T}_d \to \mathcal{U}(2)\times \mathcal{U}(2)$ corresponding to the symplectic action of the subgroup $U(2)\times U(2) \subset U_d(2,2)$ on $\mathbb{T}_d$. Recall here that $g=\left(\begin{array}{cc} A & B \\ C & D \end{array}\right) \in U_d(2,2)$ acts on $\left(\begin{array}{c} \eta \\ \xi \end{array}\right) \in \mathbb{T}_d$  by 
\begin{equation}
\left(\begin{array}{c} \eta \\ \xi \end{array}\right) \mapsto \left(\begin{array}{cc} A & B \\ C & D \end{array}\right)\left(\begin{array}{c} \eta \\ \xi \end{array}\right),
\end{equation}
where $g\in U(2)\times U(2)$ iff $B=C=0$.

Identifying the Lie algebra $\mathcal{U}(2)$ of $U(2)$ with $i H(2)$ and thus, $\mathcal{U}(2) \times \mathcal{U}(2) $ with $iH(2) \times iH(2)$, we write the momentum map $\textbf{D}_d$ as follows 
\begin{equation}
\textbf{D}_d (\eta, \xi, \eta^+ , \xi^+ ) = (iI(\eta, \eta^+), i J(\xi, \xi^+)), 
\end{equation}
where 
\begin{equation}\label{defIJ}
I(\eta, \eta^+) := \eta\eta^+ \qquad\mbox{ and } \qquad J(\xi, \xi^+) := \xi \xi^+ . 
\end{equation}
Using Pauli matrices defined in (\ref{Pauli})
as a basis of $H(2)$ one obtains 
\begin{equation}\label{basisIJ}
I = \left(\begin{array}{cc}
I_0 + I_3 & I_1+iI_2 \\
I_1-iI_2 & I_0 -I_3 
\end{array}\right)\quad 
\mbox{ and }\quad
J = \left(\begin{array}{cc}
J_0 + J_3 & J_1+iJ_2 \\
J_1-iJ_2 & J_0 -J_3 
\end{array}\right) ,
\end{equation}
where 
\begin{equation}\label{IJ}
I_\mu = \frac{1}{2} \Tr( \sigma_\mu I) = \frac{1}{2} \eta^+\sigma_\mu \eta\quad \mbox{ and } \quad J_\nu = \frac{1}{2} \Tr( \sigma_\nu J) = \frac{1}{2} \xi^+\sigma_\nu \xi
\end{equation}
for $\mu , \nu = 0,1,2,3$. The system of the functions $I_\mu, J_\nu\in C^\infty(\mathbb{T}_d,\mathbb{R})$ is closed with respect to the Poisson bracket (\ref{pb3}), i.e. 
\begin{eqnarray}
\label{pbi}
\{I_k , I_l \} = - \epsilon_{klm} I_m ,  & \quad\qquad \{I_0 , I_k \} =0,\nonumber\\
\ \\
\{J_k , J_l \} = \epsilon_{klm} J_m , & \quad\qquad \{J_0 , J_k \}=0\nonumber
\end{eqnarray}
and
\begin{equation}\label{zerobracket}
 \{I_\mu , J_\nu \}=0 
\end{equation}
for $k,l,m =1,2,3$ and $\mu, \nu = 0,1,2,3$. The above relations are a consequence of the fact that $\textbf{D}_d: (\mathbb{T}_d, \de\gamma_d)\to (\mathcal{U}(2)\times \mathcal{U}(2), \{\cdot , \cdot \}_{L-P})$ is a Poisson map. It follows from (\ref{pbi})-(\ref{zerobracket}) that the Poisson bracket of the functions $F,G\in C^\infty(\mathcal{U}(2)\times \mathcal{U}(2),\mathbb{R})$ is 
\begin{equation}
\{F, G\}_{_{L-P}} = -\epsilon_{klm} I_k\frac{\partial F}{\partial I_l}\frac{\partial G}{\partial I_m}+
 \epsilon_{klm} J_k\frac{\partial F}{\partial J_l}\frac{\partial G}{\partial J_m}
\end{equation}
and 
\be \{I_0,F\}=0=\{J_0,F\}\ee
for any function F, i.e. $I_0$ and $J_0$ are Casimirs of the Lie-Poisson space $(\mathcal{U}(2)\times \mathcal{U}(2), \{\cdot , \cdot \}_{_{L-P}})$.

From (\ref{defIJ}) and (\ref{basisIJ}) one obtains 
\begin{equation}\label{13}
|\eta_1|^2 = I_0+I_3, \mbox{ } |\eta_2|^2 = I_0 - I_3, \mbox{ } |\xi_1|^2 = J_0+J_3, \mbox{ } |\xi_2|^2 = J_0 -J_3, 
\end{equation}
\be\label{I1I2}
\eta_1\ol\eta_2   = I_1 +iI_2 ,\ee
\be\label{14}
\xi_1\ol\xi_2  = J_1 +iJ_2. 
\ee
Using (\ref{13})-(\ref{14}), we express  the Hamiltonian (\ref{H1}) in the coordinates  $({I}_\nu, {J}_\mu)$
\be \label{Hr} \tilde{H}= \tilde{H}_0(I_0,J_0,I_3,J_3)+\tilde{G}_0(I_0,J_0,I_3,J_3)((I_1+iI_2)^k(J_1+iJ_2)^l+(I_1-iI_2)^k(J_1-iJ_2)^l),\ee
where the functions $\tilde{H}_0$ and $\tilde{G}_0$ are defined by substituting  (\ref{13}) into $H_0$ and $G_0$, respectively.

The Hamilton equations on  $\mathcal{U}(2)\times \mathcal{U}(2)$ defined by the Hamiltonian (\ref{Hr}) are
\begin{eqnarray}\label{217}
\frac{\de}{\de t} {I_0}  = 0,\qquad&\quad &\qquad \frac{\de}{\de t} {I_k}=- \epsilon_{klm}I_l  \frac{\partial \tilde{H}}{\partial {I_m}},\nonumber\\
\ \\
\frac{\de}{\de t} {J_0}  = 0,\qquad&\quad &\qquad \frac{\de}{\de t} {J_k}=- \epsilon_{klm}J_l  \frac{\partial \tilde{H}}{\partial {J_m}}.\nonumber\end{eqnarray}

 Let us note that solving equations (\ref{Hequa1}) we also obtain the solution of  (\ref{217}). However, the opposite statement is not true. Therefore, let us integrate the Hamilton equations (\ref{Hequa1}). For this reason we use polar coordinates on $\mathbb{T}_d$: 
\begin{eqnarray}\label{Heqpolar1}
\eta_1=|\eta_1|e^{\frac{i}{2}(\varphi_0+\varphi_3)}\qquad&\ & \qquad \eta_2=|\eta_2|e^{\frac{i}{2}(\varphi_0-\varphi_3)}\nonumber\\
\ \\
\xi_1=|\xi_1|e^{-\frac{i}{2}(\psi_0+\psi_3)}\qquad&\ & \qquad \xi_2=|\xi_2|e^{-\frac{i}{2}(\psi_0-\psi_3)}\nonumber
\end{eqnarray}

Then, using  (\ref{13}) and (\ref{Heqpolar1}) we obtain 
\be \gamma_d=-i\de(I_0-J_0)+I_0\de\varphi_0+J_0\psi_0+I_3\de\varphi_3+J_3\de\psi_3=\qquad\ee
$$=-i\de(I_0-J_0)+I_0\de\varphi_0+J_0\psi_0+I_3' \de\varphi_3' +J_3' \de\psi_3' ,$$
where 
\be\label{linearmap}\left(\begin{array}{c}I_3'\\ J_3'\end{array}\right)=\frac{1}{k^2+l^2}\left(\begin{array}{cc}k&-l\\ l&k\end{array}\right)\left(\begin{array}{c}I_3\\ J_3\end{array}\right),\qquad
\left(\begin{array}{c}\varphi_3 '\\ \psi_3'\end{array}\right)=\left(\begin{array}{cc}k&-l\\ l&k\end{array}\right)\left(\begin{array}{c}\varphi_3\\ \psi_3\end{array}\right). \ee

The Hamiltonian (\ref{H1}) in the canonical coordinates $(I_0,J_0,I_3',J_3',\varphi_0,\psi_0,\varphi_3', \psi_3')$ assumes the following form 
\be \label{tildeH}\tilde{H}'=\tilde{H}_0'(I_0,J_0,I_3',J_3')+\tilde{G}_0'(I_0,J_0,I_3',J_3')\cos \varphi_3',\ee
where 
\be \tilde{H}_0'(I_0,J_0,I_3',J_3'):=\tilde{H}_0(J_0,J_0, kI_3'+lJ_3', -lI_3'+kJ_3')\ee
and
\be \tilde{G}_0'(I_0,J_0,I_3',J_3'):=\qquad\qquad\qquad\ee
$$=2\tilde{G}_0(J_0,J_0, kI_3'+lJ_3', -lI_3'+kJ_3')\cdot \left[I_0^2-(kI_3'+lJ_3')^2\right]^{\frac{|k|}{2}}\left[J_0^2-(-lI_3'+kJ_3')^2\right]^{\frac{|l|}{2}}.$$

The Poisson bracket (\ref{pb3}) in these canonical coordinates is given by 
\be\label{pbpolar'}
\{f,g\}=\frac{\partial f}{\partial I_0}\frac{\partial g}{\partial \varphi_0}+
\frac{\partial f}{\partial J_0}\frac{\partial g}{\partial \psi_0}+
\frac{\partial f}{\partial I_3'}\frac{\partial g}{\partial \varphi_3'}+
\frac{\partial f}{\partial J_3'}\frac{\partial g}{\partial \psi_3'}
\ee
$$-\frac{\partial f}{\partial \varphi_0}\frac{\partial g}{\partial I _0}-
\frac{\partial f}{\partial \psi_0}\frac{\partial g}{\partial J_0}-
\frac{\partial f}{\partial \varphi_3'}\frac{\partial g}{\partial I_3'}-
\frac{\partial f}{\partial \psi_3'}\frac{\partial g}{\partial J _3'}.$$
So, the Hamilton equations (\ref{Hequa1})  are equivalent to the Hamilton equations
\be\label{H5I3'} \frac{\de}{\de t}I_0=\frac{\de}{\de t}J_0=\frac{\de}{\de t}J_3'=0,\qquad
\frac{\de}{\de t}I_3'=\tilde{G}_0'(I_0,J_0,I_3', J_3')\sin \varphi_3',\ee
\begin{eqnarray} \label{H6} \frac{\de}{\de t}\varphi_0=\frac{\partial \tilde{H}'}{\partial I_0}, &\  \quad &  
\frac{\de}{\de t}\psi_3'=\frac{\partial \tilde{H}'}{\partial J_3'},\nonumber\\
\ \\
\frac{\de}{\de t}\psi_0=\frac{\partial \tilde{H}'}{\partial J_0},& \ \quad & 
\frac{\de}{\de t}\varphi_3'=\frac{\partial \tilde{H}'}{\partial I_3'},\nonumber\end{eqnarray}
defined by the Hamiltonian (\ref{tildeH}).

We see from (\ref{H5I3'}) that $I_0$,  $J_0$ and  $J_3'$ are integrals of motion. Hence, from (\ref{tildeH}) and (\ref{H5I3'}) one obtains
\be\label{dI3'} \left(\frac{\de I_3'}{\de t}(t)\right)^2=(\tilde{G}_0'(I_0,J_0,I_3'(t),J_3'))^2- (\tilde {H}'-\tilde{H}_0'(I_0,J_0,I_3'(t),J_3'))^2,\ee
where for  the integral of motion $I_0$,  $J_0$, $J_3'$ and  $\tilde {H}'$    we substituted  some constants. So, one  integrates (\ref{dI3'}) by quadratures
\be\label{calka}\pm\int \frac{\de I_3'}{\sqrt{(\tilde{G}_0'(I_0,J_0,I_3'(t),J_3'))^2- (\tilde {H}'-\tilde{H}_0'(I_0,J_0,I_3'(t),J_3'))^2}}=t+t_0.\ee

Having obtained $I_3'(t)$ and remembering that $I_0$,  $J_0$,  $J_3'$ and $\tilde {H}'$ are integrals of motion for the Hamiltonian (\ref{tildeH}) we find  using (\ref{tildeH}) and (\ref{H6}) the following system of equations
\be\label{a} \frac{\de}{\de t}\varphi_0(t)=\frac{\partial \tilde{H}_0'}{\partial I_0}(I_0,J_0,I_3'(t),J_3')+\qquad\qquad\qquad\qquad\qquad\qquad\ee
$$ +\frac{\partial \tilde{G}_0'}{\partial I_0}(I_0,J_0,I_3'(t),J_3')\cdot\frac{\tilde {H}'-\tilde{H}_0'(I_0,J_0,I_3'(t),J_3')}{\tilde{G}_0'(I_0,J_0,I_3'(t),J_3')}$$
\be\label{a2} \frac{\de}{\de t}\psi_0(t)=\frac{\partial \tilde{H}_0'}{\partial J_0}(I_0,J_0,I_3'(t),J_3')+\qquad\qquad\qquad\qquad\qquad\qquad\ee
$$ +\frac{\partial \tilde{G}_0'}{\partial J_0}(I_0,J_0,I_3'(t),J_3')\cdot\frac{\tilde {H}'-\tilde{H}_0'(I_0,J_0,I_3'(t),J_3')}{\tilde{G}_0'(I_0,J_0,I_3'(t),J_3')}$$
\be\label{a3} \frac{\de}{\de t}\varphi_3'(t)=\frac{\partial \tilde{H}_0'}{\partial I_3'}(I_0,J_0,I_3'(t),J_3')+\qquad\qquad\qquad\qquad\qquad\qquad\ee
$$ +\frac{\partial \tilde{G}_0'}{\partial I_3'}(I_0,J_0,I_3'(t),J_3')\cdot\frac{\tilde {H}'-\tilde{H}_0'(I_0,J_0,I_3'(t),J_3')}{\tilde{G}_0'(I_0,J_0,I_3'(t),J_3')}$$
\be\label{a4} \frac{\de}{\de t}\psi_3'(t)=\frac{\partial \tilde{H}_0'}{\partial J_3'}(I_0,J_0,I_3'(t),J_3')+\qquad\qquad\qquad\qquad\qquad\qquad\ee
$$ +\frac{\partial \tilde{G}_0'}{\partial J_3'}(I_0,J_0,I_3'(t),J_3')\cdot\frac{\tilde {H}'-\tilde{H}_0'(I_0,J_0,I_3'(t),J_3')}{\tilde{G}_0'(I_0,J_0,I_3'(t),J_3')}$$
for the functions $\varphi_0(t),\  \psi_0(t), \ \varphi_3'(t)$ and  $\psi_3'(t)$. The   right-hand sides of the above equations depend on $I'_3(t)$ obtained from (\ref{calka}) and on a choice of the constants of motion   $I_0$,  $J_0$,  $J_3'$ and $\tilde {H}'$  only. So, they are integrated by quadratures as well as the equation (\ref{dI3'}).

Having obtained the solution   $(I_0,J_0,I_3'(t),J_3',\varphi_0(t), \psi_0(t), \varphi_3'(t), \psi_3'(t))$ of the Hamilton  equations (\ref{H6})-(\ref{dI3'}) we find the solution $(\eta(t),\xi(t))$ of the Hamilton equation  (\ref{Hequa1})  by using the dependences (\ref{13}), (\ref{Heqpolar1}) and (\ref{linearmap}). The solution $(I_\mu(t), J_\nu(t))$ of Hamilton equations  (\ref{217}) one obtains substituting $(\eta(t),\xi(t))$
into (\ref{IJ}).

In Section \ref{sec:ex} we will  present some  examples for which the solution of the equation (\ref{Hequa1}) will be given explicitly.

		\section{Perturbed Kepler system}\label{sec:Kepler}
In this section we will show that the Hamiltonian systems of four non-linear one-dimensional oscillators investigated in the previous section  after reduction to the phase space of  null twistors give a family of  integrable perturbed Kepler systems.

Having the above in view we note that the momentum map $\textbf{J}:\mathbb{T}\to \mathcal{U}(2,2)$ maps the space $\mathbb{T}^0:=\{v\in \mathbb{T}:\ \langle v,v\rangle=0\ and\  v\not=0\}$ of null twistors  on the $Ad(U(2,2))$-orbit $\mathcal{N}^{10}$ of nilpotent elements of $\mathcal{U}(2,2)$ of rank one, see (\ref{momentummap0}) and (\ref{property0}). 

Let us define  submanifolds $\mathcal{O}_d^{10}\subset U(2)\times iH(2)\cong T^*U(2)$ and $\mathcal{O}_a^{10}\subset H(2)\times H(2)$ by
		\be \mathcal{O}_d^{10}:=U(2)\times i\left\{\rho \in H(2):\ \det\rho=0\ {\rm and}\ \Tr\rho>0\right\}\ee
		and by
			\be \mathcal{O}_a^{10}:=H(2)\times \left\{X\in H(2):\ \det X=0\  {\rm and}\ \Tr X>0 \right\},\ee
		respectively. Restricting  the morphisms of diagram (\ref{diagram}) to the above submanifolds and the spaces of null twistors we obtain  
		
		 \unitlength=5mm \begin{equation}\label{diagram2}\begin{picture}(11,4.6)
    \put(-2,4){\makebox(0,0){$\mathcal{O}_d^{10}$}}
    \put(5,4){\makebox(0,0){$\mathcal{N}_d^{10}$}}
    \put(12.5,4){\makebox(0,0){$\mathbb{T}^0_d$}}
    \put(12.5,0){\makebox(0,0){$\mathbb{T}^0_a,$}}
    \put(5,0){\makebox(0,0){$\mathcal{N}_a^{10}$}}
    \put(-2,0){\makebox(0,0){$\mathcal{O}_a^{10}$}}

    \put(-2,1){\vector(0,1){2}}
    \put(5,1){\vector(0,1){2}}
    \put(12,1){\vector(0,1){2}}
    \put(11,4){\vector(-1,0){4}}
    \put(11,0){\vector(-1,0){4}}
 \put(0.2,0){\vector(1,0){3}}
 \put(0,4){\vector(1,0){3}}
    \put(1.5,4.4){\makebox(0,0){$\textbf{I}_d$}}
     \put(8.5,4.4){\makebox(0,0){$\textbf{J}_d$}}
    \put(8.5,0.5){\makebox(0,0){$\textbf{J}_a$}}
\put(1.5,0.5){\makebox(0,0){$\textbf{I}_a$}}
  \put(-1,2){\makebox(0,0){$T^*_{\mathcal{C}}$}}
    \put(6.3,2){\makebox(0,0){$Ad_\mathcal{C}$}}
\put(13,2){\makebox(0,0){$ \mathcal{C}$}}
\put(-2.20,1){\makebox(0,0){$\cup$}}
\put(0,0.2){\makebox(0,0){$\subset$}}

    \end{picture}\end{equation}
		where one has
	$$ \textbf{I}_d(\mathcal{O}_d^{10})=\mathcal{N}_d^{10}=\textbf{J}_d(\mathbb{T}_d^{0}) \qquad {\rm and}\qquad \textbf{J}_a(\mathbb{T}_a^{0})=\mathcal{N}_a^{10}.$$

Let us describe the level sets  of the horizontal morphisms in the diagram (\ref{diagram2}). 
\begin{enumerate}[(i)]
\item For  surjective submersions $\textbf{J}_d:\mathbb{T}_d^0\to  \mathcal{N}_d^{10}$ and $\textbf{J}_a:\mathbb{T}_a^0\to  \mathcal{N}_a^{10}$ these level sets  are orbits of the group $U(1)$ which acts on $\mathbb{T}$ by 
\be\label{UT} U(1)\times\mathbb{T} \ni(e^{it},v)\mapsto e^{it}v\in \mathbb{T}.\ee 
Note that the action (\ref{UT}) is the Hamiltonian flow on the twistor space $(\mathbb{T},\de\gamma)$ which is  generated by the function $\textbf{L}_{\mathfrak{X}}\circ \textbf{J}:\mathbb{T}\to \mathbb{R}$, where  $\mathfrak{X}=-i\phi\in \mathcal{U}(2,2)$, see (\ref{momentummap0}) and (\ref{lJ}).
\item For the momentum map $\textbf{I}_d:\mathcal{O}_d^{10}\to \mathcal{N}_d^{10}$ one has 
\be\label{rel1} (Z',\rho')\in \textbf{I}_d^{-1}(\textbf{I}_d(Z,\rho)) \quad {\rm iff}\quad \rho'=\rho\quad{\rm and }\quad Z'\rho'=Z\rho.\ee
\item For $\textbf{I}_a:H(2)\times H(2)\to \mathcal{N}^{10}_a$ one has 
\be\label{rel2} (Y',X')\in \textbf{I}_a^{-1}(\textbf{I}_a(Y,X)) \quad {\rm iff}\quad X'=X \quad{\rm and }\quad Y'- Y=\lambda\tilde X,\ee
where $\lambda\in \mathbb{R}$ and $\tilde X=X_0E-\vec{X}\vec\sigma$ if $X=X_0E+\vec{X}\vec\sigma$.
\end{enumerate}

Now we show that the vertical morphisms of (\ref{diagram2}) preserve the equivalence relations defined in $(i)$, $(ii)$ and $(iii)$. The map $\mathcal{C}:\mathbb{T}_a\to \mathbb{T}_d$ is a linear map, so, it commutes with the action (\ref{UT}) of $ U(1)$ on $\mathbb{T}_a$ and $\mathbb{T}_d $ and thus on $\mathbb{T}_a^0$ and $\mathbb{T}_d ^0$. 
Using (\ref{z}) and (\ref{rho}) we find that $(Z',\rho')\sim (Z,\rho)$, where $(Z',\rho')=T_\mathcal{C}^*(Y',X')$ and $(Z,\rho)=T_\mathcal{C}^*(Y,X)$, if and only if 
\be\label{XY}\begin{array}{l}
		X'=X\\
		\  \\
		X(Y'-Y)+(Y'-Y)X=0.\end{array}\ee
		Remembering that $X=\zeta\zeta^+$ we obtain from (\ref{XY}) that		
	\be\label{49} \zeta^+(Y'-Y)\zeta=0\ee
		and 
		\be\label{410} (Y'-Y)\zeta=\frac{ \zeta^+(Y'-Y)\zeta}{\zeta^+\zeta}\zeta=0.\ee
	Now, from (\ref{XY}), (\ref{49}) and (\ref{410}) we find that $X'=X$ and $ (Y'-Y)X=0$, i.e. $(Y',X')\sim (Y,X)$ in sense of (\ref{rel2}).

		\bigskip

		Below we will use the following notation $\tilde M$, $\tilde N$ for the quotient manifolds $M/_\sim$,  $N/_\sim$  with respect to some equivalence relations. Consistently with the above, by $\tilde \phi:\tilde M\to\tilde N$ we will denote the quotient manifolds map defined by the the map $\phi:M\to N$, which by definition preserves these equivalence relations.
		
		Applying this notation, after quotienting the diagram (\ref{diagram2}) by the equivalence relations described in $(i)$, $(ii)$ and $(iii)$, we formulate the following proposition.
		
	\begin{prop}\label{prop1}
	The objects of the diagram 
	\unitlength=5mm \begin{equation}\label{diagram3}\begin{picture}(11,4.6)
    \put(-2,4){\makebox(0,0){$\tilde{\mathcal{O}}_d^{10}$}}
    \put(5,4){\makebox(0,0){$\mathcal{N}_d^{10}$}}
    \put(12.5,4){\makebox(0,0){$\tilde{\mathbb{T}}^0_d$}}
    \put(12.5,0){\makebox(0,0){$\tilde{\mathbb{T}}^0_a,$}}
    \put(5,0){\makebox(0,0){$\mathcal{N}_a^{10}$}}
    \put(-2,0){\makebox(0,0){$\tilde{\mathcal{O}}_a^{10}$}}

    \put(-2,1){\vector(0,1){2}}
    \put(5,1){\vector(0,1){2}}
    \put(12,1){\vector(0,1){2}}
    \put(11,4){\vector(-1,0){4}}
    \put(11,0){\vector(-1,0){4}}
 \put(0.4,0){\vector(1,0){3}}
 \put(0,4){\vector(1,0){3}}
    \put(1.5,4.5){\makebox(0,0){$\tilde{\textbf{I}_d}$}}
     \put(8.5,4.5){\makebox(0,0){$\tilde{\textbf{J}_d}$}}
    \put(8.5,0.5){\makebox(0,0){$\tilde{\textbf{J}_a}$}}
\put(1.5,0.5){\makebox(0,0){$\tilde{\textbf{I}_a}$}}
  \put(-1,2){\makebox(0,0){$\tilde{T}^*_{\mathcal{C}}$}}
    \put(6.3,2){\makebox(0,0){$\widetilde{Ad_\mathcal{C}}$}}
\put(13,2){\makebox(0,0){$ \tilde{\mathcal{C}}$}}
\put(-2.20,1){\makebox(0,0){$\cup$}}
\put(0.2,0.2){\makebox(0,0){$\subset$}}

    \end{picture}\end{equation}
		are symplectic manifolds.  The arrows of this diagram define the symplectomorphisms between these symplectic manifolds. The maps $\tilde{T}^*_{\mathcal{C}}: \tilde{\mathcal{O}}_a^{10}\hookrightarrow \tilde{\mathcal{O}}_d^{10}$ and $ \tilde{\textbf{I}_a}: \tilde{\mathcal{O}}_a^{10}\hookrightarrow \mathcal{N}_a^{10}$ are symplectic embedings of $\tilde{\mathcal{O}}_a^{10}$ onto open and dense subsets of $\tilde{\mathcal{O}}_d^{10}$ and $\mathcal{N}_a^{10}$, respectively. The other maps in (\ref{diagram3}) are symplectic diffeomorphisms.
		
		\end{prop}
		
		\bigskip
	
	In order to obtain the diffeomorphism  $\tilde{\textbf{J}}^{-1}_d\circ \tilde{\textbf{I}}_d:\tilde{\mathcal{O}}_d^{10}\stackrel{\sim }{\to}\tilde{\mathbb{T}}^0_d$ explicitly, let us note  that if $(Z,\rho) \in \mathcal{O}_d^{10}$ then $\rho=\xi\xi^+$ for some spinor $\xi\in\dot {\mathbb{C}}^2$, where $\dot {\mathbb{C}}^2:={\mathbb{C}}^2\setminus\{0\}$, which is defined by  $\rho$ up to a factor $\lambda\in U(1)\subset \mathbb{C}\setminus\{0\}$. Thus for $[(Z,\rho)]\in \tilde{\mathcal{O}}_d^{10}$ we have 
	\be (\tilde{\textbf{J}}^{-1}_d\circ \tilde{\textbf{I}}_d)([(Z,\rho)])=\left[\left(\begin{array}{c}Z\xi\\ \xi\end{array}\right)\right].\ee
	 Note that from $(Z',\rho ')\in [(Z,\rho)]$ follows that $Z^+Z'\xi=\xi$, i.e. $Z\in U(2)$ is determined by $\left[\left(\begin{array}{c}\eta\\ \xi\end{array}\right)\right]=\left[\left(\begin{array}{c}Z\xi\\ \xi\end{array}\right)\right]\in \mathbb{T}_d^0$ up to the stabilizer $U(2)_{[\xi]}\subset U(2)$ of $[\xi]\in \dot {\mathbb{C}}^2/_{U(1)}$. We can remove this ambiguity  assuming that $Z\in iSU(2)$. Hence, we find that the symplectic manifolds of the upper row in the diagram (\ref{diagram3}) and $\tilde {\mathbb{T}}_a^0$ are diffeomorphic
to 
\be iSU(2)\times 	 \dot {\mathbb{C}}^2/_{U(1)}\cong  \mathbb{S}^3\times \dot {\mathbb{R}}^3,\ee
where $\dot {\mathbb{R}}^3:={\mathbb{R}}^3\setminus\{0\}$.
Recall here that $g\in SU(2)$ iff $g=\left(\begin{array}{rc}a&b\\ -\ol{b}&\ol{a}\end{array}\right)$, where  $a,b\in \mathbb{C}$ satisfy $|a|^2+|b|^2=1$. 

Let us note  that $\tilde{T}_\mathcal{C}^*(\tilde{\mathcal{O}}_a^{10})=\tilde{\dot{\mathcal{O}}}_d^{10}$ and  $(\tilde{\textbf{J}}^{-1}_a\circ \tilde{\textbf{I}}_a)(\tilde{\mathcal{O}}_a^{10})=\tilde{\dot{\mathbb{T}}}^0_a$ where
\be \dot{\mathcal{O}}_d^{10}:=\{(Z,\rho)\in {\mathcal{O}}_d^{10}:\quad \det(iZ+\sigma_0)\not=0\}\ee
and
\be {\dot{\mathbb{T}}}^0_a:= \left\{\left(\begin{array}{l}\vartheta\\ \zeta\end{array}\right)\in \mathbb{T}^0_a:\quad \zeta\not=0\right\},\ee
respectively. For finding  symplectic difeomorphism $\tilde{\textbf{J}}^{-1}_a\circ \tilde{\textbf{I}}_a:\tilde{\mathcal{O}}_a^{10}\stackrel{\sim }{\to}\tilde{\dot{\mathbb{T}}}^0_a$ and its inverse  $\tilde{\textbf{I}}_a^{-1}\circ \tilde{\textbf{J}}_a:\tilde{\dot{\mathbb{T}}}^0_a\stackrel{\sim }{\to}\tilde{\mathcal{O}}_a^{10}$ explicitly we use the equality $\textbf{J}_a(\vartheta,\zeta)=\textbf{I}_a(X,Y)$, which is equivalent to 
\be\label{XYa} X=\zeta\zeta^+,\qquad {\rm and}\qquad Y\zeta=\vartheta.\ee
From (\ref{XYa}) we have 
\be\label{JIa} (\tilde{\textbf{J}}^{-1}_a\circ \tilde{\textbf{I}}_a)([(Y,X)])=\left[\left(\begin{array}{l}Y\zeta\\ \zeta\end{array}\right)\right].\ee
One easily sees from (\ref{rel2}) and (\ref{XYa}) that this formula  does not  depend on the choice of $(X', Y')\in [(Y,X)]$.
The map inverse to (\ref{JIa}) is given by 
\be\label{IiJ} (\tilde{\textbf{I}}_a^{-1}\circ \tilde{\textbf{J}}_a)\left(\left[\left(\begin{array}{l}\vartheta\\ \zeta\end{array}\right)\right]\right)=
[(Y(\vartheta,\zeta,\vartheta^+,\zeta^+),X(\vartheta,\zeta,\vartheta^+,\zeta^+)],\ee
where \be\label{Ya} Y(\vartheta,\zeta,\vartheta^+,\zeta^+):=\frac{1}{\zeta^+\zeta}\left[\zeta \vartheta^++\vartheta\zeta^+-\frac{1}{2}(\vartheta^+\zeta+\zeta^+\vartheta)\sigma_0\right]\ee
\be\label{Xa} X(\vartheta,\zeta,\vartheta^+,\zeta^+):=\zeta\zeta^+.\qquad\qquad\qquad\qquad\qquad\qquad\qquad \ee
For more information concerning the above questions see \cite{AO}. Here we note that for $Y(\vartheta,\zeta,\vartheta^+,\zeta^+)$ defined in (\ref{Ya})  one has $\Tr Y(\vartheta,\zeta,\vartheta^+,\zeta^+)=0$. Having in mind that the equivalence class $[(Y,X)]\in \mathcal{O}^{10}_a$ has the one and only one representative $(Y,X)\in [(Y,X)]$ for which $\Tr Y=0$ we can identify $\mathcal{O}^{10}_a$ with $ \mathbb{R}^3\times \dot {\mathbb{R}}^3$ by using  decomposition 
\be (Y,X)=(y^0\sigma_0+\vec{y}\cdot \vec{\sigma},\ x^0\sigma_0+\vec{x}\cdot \vec{\sigma})\ee
on Pauli matrices, where $y^0=\frac{1}{2} TrY=0$ and $x^0=(({\vec{x}}^2)^{\frac{1}{2}}=||\vec{x}||.$ The reduced symplectic form $\widetilde{\de\pi}_a$ on $\tilde{\mathcal{O}}_a^{10}$, see (\ref{pia}), in the coordinates $(\vec{y}, \vec{x})\in \mathbb{R}^3\times \dot {\mathbb{R}}^3\cong \tilde{\mathcal{O}}_a^{10}$ is given by 
\be \widetilde{\de\pi_a}=2\de \vec{x}\wedge \de \vec{y}.\ee

Writting the symplectic diffeomorphism (\ref{IiJ}) in terms of the canonical coordinates $(\vec{y}, \vec{x})\in \mathbb{R}^3\times \dot {\mathbb{R}}^3$ and the twistor coordinates $\left(\begin{array}{l}\vartheta\\ \zeta\end{array}\right)\in \dot{\mathbb{T}}_a^0$ we see it is the Kustaanheimo-Stiefel transformation 
\be\label{KSy} \vec{y}=\frac{1}{2}\frac{1}{\zeta^+\zeta}(\vartheta^+\vec{\sigma}\zeta+\zeta^+\vec{\sigma}\vartheta),\ee
\be\label{KSx}  \vec{x}=\frac{1}{2}\zeta^+\vec{\sigma}\zeta,\qquad\qquad\qquad\qquad\ee
disscussed firstly in  \cite{stiefel}, see also \cite{kummer}. Let us summarize the above facts.
\begin{prop}\label{prop:32}
The following symplectic manifolds 
\be(\mathbb{R}^3\times \dot{\mathbb{R}} ^3, \de\vec{y}\wedge \de\vec{x})\cong (\tilde{\mathcal{O}} _a^{10},\widetilde{\de\pi}_a)\cong
(\tilde{\dot{\mathbb{T}}}_a^0, \widetilde{\de\gamma}_a)\cong (\tilde{\dot{\mathbb{T}}}_d, \widetilde{\de\gamma}_d)\ee
are symplectically diffeomorphic, where the symplectic forms  $\widetilde{\de\gamma}_d$, $\widetilde{\de\gamma}_a$ and $\widetilde{\de\pi}_a$  are obtained by the reductions to suitable submanifolds of the symplectic forms ${\de\gamma}_d$, ${\de\gamma}_a$ and ${\de\pi}_a$, defined in (\ref{gammad}), (\ref{gammaa}) and (\ref{pia}).
\end{prop}

\bigskip

Let us define $R,M\in H(2)$ by 
\be\label{R1} R:=I+J\qquad {\rm and}\qquad M:=J-I,\ee
where the  matrix coordinates $(I,J)\in H(2)\times H(2)$ were defined in (\ref{basisIJ}). We observe that $R$ and $M$ in the coordinates $(Y, X)\in H(2)\times H(2)$  are given by
\be\label{R2} R=X+XYX\qquad {\rm and}\qquad M=i[X,Y],\ee
  Expanding  (\ref{R2}) in Pauli matrices, i.e. $R=R_\mu\sigma_\mu$ and $M=M_\mu\sigma_\mu$, we obtain

\be  M_0=0,\qquad\qquad \vec{M}=2\vec{x}\times \vec{y},\ee
\be \label{428}  R_0=||\vec{x}||(1+{\vec{y}}^2),\qquad \vec{R}=(1-{\vec{y}}^2)\vec{x}+2(\vec{x}\cdot \vec{y})\vec{y},
\ee 
 So, $\vec{R}$ and $\vec{M}$ are the Runge-Lenz and angular momentum vectors,
respectively. 
The Hamiltonian system  of the four non-linearly interacting one-dimensional harmonic oscillators $(\mathbb{T}_d, d\gamma_d, H)$, where $H$ is defined in (\ref{H1}), possesses $I_0-J_0=\frac{1}{2}(\eta^+\eta-\xi^+\xi)$ as its integral of motion. So, one can reduce this system to a Hamiltonian system  on the symplectic manifold $(\tilde{\mathbb{T}}_d^0, \widetilde{\de\gamma_d})$. Then, using the symplectic diffeomorphisms mentioned in Proposition \ref{prop:32}, one can realize this reduced system as a Hamiltonian system on $(\mathbb{R}^3\times \dot{\mathbb{R}} ^3, 2\de\vec{y}\wedge \de\vec{x})$, where the reduced  Hamiltonian   is given by 
\be\label{HPK} H_{PK}:= \tilde{H}_0(\frac{1}{2}R_0,\frac{1}{2}R_0,I_3, J_3)+\qquad\qquad\qquad\qquad\qquad\qquad\qquad\ee
$$+\tilde{G}_0(\frac{1}{2}R_0,\frac{1}{2}R_0,I_3, J_3)((I_1+iI_2)^k(J_1+iJ_2)^l+(I_1-iI_2)^k(J_1+iJ_2)^l),$$
where 
\be \label{3I}{I_k}=\frac{1}{2}(1-{\vec{y}}^2){x_k}+(\vec{x}\cdot \vec{y}){y_k}-\epsilon_{klm}x_ly_m,\ee
\be \label{3J}{J_k}=\frac{1}{2}(1-{\vec{y}}^2){x_k}+(\vec{x}\cdot \vec{y}){y_k}+\epsilon_{klm}x_ly_m,\qquad\ee
and for $R_0$ see (\ref{428}).
One obtains $H_{PK}$ substituting to $\tilde{H}$, see (\ref{Hr}), $I_0=\frac{1}{2}R_0$ and $J_0=\frac{1}{2}R_0$. Note here that $\left(\begin{array}{c}\eta\\ \xi\end{array}\right)\in \mathbb{T}^0_d$ iff $M_0=I_0-J_0=0$.

In particular case when $\tilde{H}(\frac{1}{2}R_0,\frac{1}{2}R_0,I_3, J_3)=R_0$ and  $\tilde{G}(\frac{1}{2}R_0,\frac{1}{2}R_0,I_3, J_3)\equiv 0$ the Hamiltonian (\ref{HPK}) gives the Hamiltonian for Kepler system
\be\label{HK1} H_K(\vec{x},\vec{y})= R_0=||\vec{x}||(1+{\vec{y}}^2)\ee
 written in the "fictitious time" $t$ related to the real time $t_r$ by the relation 
$$ \de t=\frac{1}{||\vec{x}||} \de t_r.$$
 For an exhaustive discussion of the time regularization problem for instance see \cite{kummer,SS}.

One obtains the solution $(\vec{y}(t),\vec{x}(t))$ of the Hamilton equations 
\begin{eqnarray}\label{HPKe}\frac{\de}{\de t}\vec{y}=-\frac{1}{2}\frac{\partial H_{PK}}{\partial\vec{x}},\nonumber\\
\ \\
\frac{\de}{\de t}\vec{x}=\frac{1}{2}\frac{\partial H_{PK}}{\partial\vec{y}}\quad\nonumber\end{eqnarray}
from the solution $(\eta(t),\xi(t))$ of (\ref{Hequa1}) by applying to $\vartheta(t)=\frac{1}{\sqrt{2}}(\eta(t)+i\xi(t))$ and $\zeta(t)=\frac{1}{\sqrt{2}}(i\eta(t)+\xi(t))$ the Kustaanheimo-Stiefel transformation defined in (\ref{KSy}) and  (\ref{KSx}).

In particular case the solution of the Hamilton equations 
\begin{eqnarray}\label{HK} \frac{\de}{\de t}\vec{y}=-\frac{1}{2}\frac{\partial H_{K}}{\partial\vec{x}}=-\frac{1}{2}(1+\vec{y}^2)\frac{\vec{x}}{||\vec{x}||},\nonumber\\
\ \\
 \frac{\de}{\de t}\vec{x}=\frac{1}{2}\frac{\partial H_{K}}{\partial\vec{y}}=||\vec{x}||\vec{y}\qquad\qquad\qquad\nonumber\end{eqnarray}
for Kepler system one easily obtains  observing that the Hamiltonian $H_K$, see (\ref{HK1}), written on $\tilde{\mathbb{T}}_d^0$ has the form
\be H_K=(\eta^+\eta+\xi^+\xi)=(\textbf{L}_\mathfrak{X}\circ \textbf{J}_d)(\eta,\xi,\eta^+,\xi^+)\ee
where $\mathfrak{X}=-i\left(\begin{array}{cc}\sigma_0&0\\ 0&-\sigma_0\end{array}\right).$ So, its Hamiltonian flow is given by 

\be\label{solK1} \left(\begin{array}{c}\eta(t)\\ \xi(t)\end{array}\right)=\left(\begin{array}{cc}e^{it}\sigma_0&0\\ 0&e^{-it}\sigma_0\end{array}\right)\left(\begin{array}{c}\eta(0)\\ \xi(0)\end{array}\right).\ee

Hence, substituting 
\be\label{solK2}\vartheta(t)=\frac{1}{\sqrt{2}}(e^{it}\eta(0)+ie^{-it}\xi(0))\ee
and 
\be\label{solK3}\zeta(t)=\frac{1}{\sqrt{2}}(ie^{it}\eta(0)+e^{-it}\xi(0))\ee
into (\ref{KSy}) and (\ref{KSx}) we obtain the solution $(\vec{y}(t),\vec{x}(t))$ of (\ref{HK}). The solutions of Hamilton equations (\ref{HPKe}) for other subcases of the Hamiltonian  (\ref{HPK}) we  will present in the next section.

Ending this section we conclude that  for small $\tilde{G}_0(\frac{1}{2}R_0,\frac{1}{2}R_0,I_3, J_3)\approx 0$ and $\tilde{H}_0(\frac{1}{2}R_0,\frac{1}{2}R_0,I_3, J_3)=R_0$ the Hamiltonian $H_{PK}$ could be  treated as  the Hamiltonian of a perturbed Kepler system.

\newpage
		\section{An example}\label{sec:ex}
		
				As an example we will consider simultaneously two cases, i.e when integers $k,l\in \mathbb{Z}$ are equal $k=1$,$\ l= 1$ or $k=1$,$\ l= -1$. We also assume $G_0 (|\eta_1|^2, |\eta_2|^2, |\xi|_1^2 , |\xi_2|^2)$  as a constant $G_0$ and put $H_0(|\eta_1|^2, |\eta_2|^2, |\xi|_1^2 , |\xi_2|^2)= |\eta_1|^2 + |\eta_2|^2 +|\xi_1|^2 +|\xi_2|^2$. Under such assumptions the Hamiltonian  (\ref{H1}) is given by 
	\be\label{exH1}
	H= |\eta_1|^2 + |\eta_2|^2 +|\xi_1|^2 +|\xi_2|^2 + G_0 \cdot (\eta_1\bar\eta_2 \xi_1 \bar\xi_2 + \bar\eta_1 \eta_2 \bar\xi_1\xi_2 )
	\ee
for  $k=1$,$\ l= 1$, and by 
\be\label{exH1-}
	H= |\eta_1|^2 + |\eta_2|^2 +|\xi_1|^2 +|\xi_2|^2 + G_0 \cdot (\eta_1\bar\eta_2 \bar\xi_1 \xi_2 + \bar\eta_1 \eta_2 \xi_1\bar\xi_2 ) 
	\ee
for $k=1$,$\ l=- 1$. 
Rewriting the above Hamiltonians in terms of the canonical coordinates $(I_0, I_0, I_3', J_3', \varphi_0, \psi_0, \varphi_3' , \psi_3')$ defined in (\ref{13}),(\ref{Heqpolar1}) and (\ref{linearmap}) one obtains 
\be\label{extildeH}
\tilde{H}'= 2(I_0 + J_0) + 2G_0 \sqrt{(I_0^2 -(I_3'\pm J_3')^2)(J_0^2 -(I_3'\mp J_3')^2)} \cos \varphi_3',
\ee
for $k=1$,$\ l=\pm 1$. In the subsequent, just like above, we will write in front of  $J_3'$ the upper sign  for $l=1$ and lower sign for $l=-1$, respectively. 

Note that the real variables $(I_0, J_0, I_3', J_3')$ must obey 
\be\begin{array}{cc}
J_0 >0 ,\quad & \frac{1}{2}(I_0 +J_0)> J_3' > -\frac{1}{2}(I_0 +J_0), \\
I_0>0 , \quad & \min \{ I_0 \mp J_3', J_0 \pm J_3' \} > I_3' > \max \{ -I_0 \mp J_3' , -J_0 \pm J_3'\} . 
\end{array} \ee
One easily sees that Hamilton equations for the Hamiltonian (\ref{extildeH}) written in the canonical coordinates $(I_0, J_0, I_3', J_3',\varphi_0,\psi_0, \varphi_3', \psi_3')$ are the following
\begin{align}
\label{exeq1}
\frac{\de}{\de t} I_0 (t) & = \frac{\de}{\de t} J_0(t)=\frac{\de}{\de t} J_3' (t)= 0, \\
\label{exdifeqi3}
\frac{\de}{\de t} I_3' (t) & = 2G_0 \sqrt{(I_0^2 -(I_3(t)'\pm J_3')^2)(J_0^2 -(I_3'(t)\mp J_3')^2)} \sin \varphi_3'(t),\\
\label{ex1}
\frac{\de}{\de t} \varphi_0 (t) & = 2 + 2G_0 I_0 \sqrt{\frac{J_0^2 -(I_3'(t)\mp J_3')^2}{I_0^2 -(I_3'(t)\pm J_3')^2}} \cos \varphi_3'(t), \\
\frac{\de}{\de t} \psi_0 (t) & = 2 + 2G_0 J_0 \sqrt{\frac{I_0^2 -(I_3'(t)\pm J_3')^2}{J_0^2 -(I_3'(t)\mp J_3')^2}} \cos \varphi_3'(t)\\
\frac{\de}{\de t} \varphi_3' (t) & = \left(\mp (I_3'(t)\mp J_3') \sqrt{\frac{I_0^2 -(I_3'(t)\pm J_3')^2}{J_0^2 -(I_3'(t)\mp J_3')^2}}-(I_3'(t) \pm J_3')\sqrt{\frac{J_0^2 -(I_3'(t)\mp J_3')^2}{I_0^2 -(I_3'(t)\pm J_3')^2}}\right) 2G_0\cos \varphi_3'(t),\\
\label{ex2}
\frac{\de}{\de t} \psi_3' (t) & = \left((I_3'(t)\mp J_3') \sqrt{\frac{I_0^2 -(I_3'(t)\pm J_3')^2}{J_0^2 -(I_3'(t)\mp J_3')^2}}\mp (I_3'(t) \pm J_3')\sqrt{\frac{J_0^2 -(I_3'(t)\mp J_3')^2}{I_0^2 -(I_3'(t)\pm J_3')^2}}\right) 2G_0\cos \varphi_3'(t). 
\end{align}
Using  (\ref{exdifeqi3}) and (\ref{extildeH}) we obtain the  differential equation 
\be\label{exdI3'}
\frac{\de}{\de t} I_3'(t) = \sqrt{4G_0^2(I_0^2 -(I_3(t)'\pm J_3')^2)(J_0^2 -(I_3'(t)\mp J_3')^2)- (\tilde{H}' -2(I_0+J_0))^2}
\ee
on the function $I_3'(t) $. After reduction to $\mathbb{T}_d^0$, what means
 $J_0 = I_0$,  and separating the variables in (\ref{exdI3'}) we find that   
\be\label{exint1}
2G_0(t-t_0) = \int_{I_3'(t_0)}^{I_3'(t)} \frac{ds}{\sqrt{(s^2 - \lambda_-^2)(s^2-\lambda_+^2)}}, 
\ee 
where $\lambda_\pm^2 := I_0^2 + J_3'^2 \pm \sqrt{2I_0^2J_3'^2 + \left(\frac{\tilde{H}'-4I_0}{2G_0}\right)^2}$. 

In order to simplify further calculations we will assume the following  initial condition $(t_0, I_3'(t_0)) = (0,0)$ for the solution $I_3'(t)$ of  (\ref{exdI3'}). Hence, for $0< I_3'(t) \leq{\lambda_-}< {\lambda_+}$,  the solution of (\ref{exdI3'}) is given by 
\be\label{solI3'}
I_3'(t) = {\lambda_-} \mbox{sn} \left( \omega t, \kappa\right) , 
\ee
where 
\be
\omega := 2G_0 {\lambda_+} \quad\mbox{ and }\quad \kappa:= {\frac{\lambda_-}{\lambda_+}}.
\ee
The function  $\mbox{sn}(\omega t,\kappa)$  is the Jacobi elliptic function called the \textit{elliptic sine}. For solution of (\ref{exint1}) and for more information about Jacobi elliptic functions see entry \textbf{3.152.7}  and Chapter \textbf{8} in \cite{ryzhik}.

Let us stress that  (\ref{solI3'}) gives the solution of Hamilton equation (\ref{exdifeqi3}) for both considered cases, i.e. $l=1$ and $l=-1$. 

Assuming  $J_3' =0$ in (\ref{ex1})-(\ref{ex2})  we obtain the equations 
\begin{align}
\label{exe1}
\frac{\de}{\de t} \varphi_0 (t) & =  2 +  I_0 \frac{\tilde{H}'- 4I_0}{I_0^2 -I_3'(t)^2}, \\
\frac{\de}{\de t} \psi_0 (t) & =  2 +  I_0 \frac{\tilde{H}'- 4I_0}{I_0^2 -I_3'(t)^2}, \\
\frac{\de}{\de t} \varphi_3' (t) & = -2 I_3'(t) \frac{\tilde{H}'- 4I_0}{I_0^2 -I_3'(t)^2},\\
\label{exe3}
\frac{\de}{\de t} \psi_3' (t) & = 0. 
\end{align}

After substituting (\ref{solI3'}), where we put $J_3'=0$,  into (\ref{exe1})-(\ref{exe3}) and  next integrating obtained in this way equations we find the solution 
\be\label{519}
\varphi_0 (t)= \varphi_0 (0)   +
 2t+\frac{\tilde{H}'-4I_0}{I_0\omega}\Pi \left(\frac{\lambda_-^2}{I_0^2};\mbox{am}(\omega t, \kappa), \kappa\right), \qquad\qquad\qquad
\ee
\be
\psi_0 (t) = \psi_0(0) +
 2t+\frac{\tilde{H}'-4I_0}{I_0\omega}\Pi \left(\frac{\lambda_-^2}{I_0^2};\mbox{am}(\omega t, \kappa), \kappa\right),\qquad\qquad\qquad\ee
\be
\psi_3' (t)  = \psi_3'(0),\qquad\qquad\qquad\qquad\qquad\qquad\qquad\qquad \qquad\qquad\qquad\qquad \qquad 
\ee
\begin{multline}\label{522}
\varphi_3' (t) = \varphi_3' (0) + \frac{\tilde{H}' - 4I_0}{4G_0\lambda_-\lambda_+\sqrt{(  \lambda_-^2-I_0^2)( \lambda_+^2-I_0^2)}} \times \\
\ln 
\left(\frac{\sqrt{\left(\lambda_-^2-I_0^2\right)\left(1 -\kappa^2\ \mbox{sn}^2 (\omega t,\kappa)\right)}- 
\kappa\sqrt{\left(\lambda_+^2-I_0^2\right)\left(1-\mbox{sn}^2(\omega t,\kappa)\right)}}
{\sqrt{\left(\lambda_-^2-I_0^2\right)\left(1 -\kappa^2\ \mbox{sn}^2 (\omega t,\kappa)\right)}+ 
\kappa\sqrt{\left(\lambda_+^2-I_0^2\right)\left(1-\mbox{sn}^2(\omega t,\kappa)\right)}}\right) ,
\end{multline}
of (\ref{exe1})-(\ref{exe3}), where $ \mbox{am}(\omega t,\kappa) $ is the Jacobi amplitude function and $\Pi(n; \omega t,\kappa)$ is the incomplete elliptic integral of the third kind, see \cite{ryzhik} for their definitions and some properties. 

Let us note here that $\psi_0(t)=\varphi_0(t)+\psi_0(0)-\varphi_0(0)$. In the below we will use the following notation: $\psi_3'(0)=:\Delta_1$ and $\psi_0(0)-\varphi_0(0)=:\Delta_2$.

Now,  we obtain the time dependence of the complex variables $\eta_1(t), \eta_2(t),$ $ \xi_1(t) , \xi_2(t)$. For these reason we apply the transformations (\ref{13}), (\ref{Heqpolar1}) and (\ref{linearmap}) to the functions $I_3'(t),$ $\varphi_0(t),$  $\psi_0(t)=\varphi_0(t)+\Delta_2$ , $\varphi_3'(t)$ and to the constants $I_0=J_0$, $J_3'=0$. As a results we obtain
\begin{align}
\nonumber
\eta_1 (t) & = \sqrt{I_0 + I_3'(t)}e^{\frac{i}{2} ( \varphi_0 (t) +\frac{1}{2}(\varphi_3'(t)+\Delta_1))}, \\
\nonumber
\eta_2 (t) & = \sqrt{I_0 - I_3'(t)}e^{\frac{i}{2} (\varphi_0(t) - \frac{1}{2}(\varphi_3' (t) + \Delta_1))}, \\
\label{eta11}
\xi_1 (t) & = e^{-\frac{i}{2}(\Delta_2 + \Delta_1)}\bar \eta_2 (t),\\
\nonumber
\xi_2 (t) & = e^{-\frac{i}{2}(\Delta_2 - \Delta_1)}\bar \eta_1 (t)
\end{align}
for the case $l=1$ and 
\begin{align}\nonumber
\eta_1 (t) & = \sqrt{I_0 + I_3'(t)}e^{\frac{i}{2} ( \varphi_0 (t) +\frac{1}{2}(\varphi_3'(t)- \Delta_1))}, \\
\nonumber
\eta_2 (t) & = \sqrt{I_0 - I_3'(t)}e^{\frac{i}{2} (\varphi_0(t) - \frac{1}{2}(\varphi_3' (t) - \Delta_1))}, \\
\label{eta1-1}
\xi_1 (t) & = e^{-\frac{i}{2}(\Delta_2 + \Delta_1)}\bar \eta_1 (t),\\
\nonumber
\xi_2 (t) & = e^{-\frac{i}{2}(\Delta_2 - \Delta_1)}\bar \eta_2 (t)
\end{align}
for the case $l=-1$.

In order to find explicitly  the Hamiltonian flow
\be \mathbb{R}\ni t\mapsto 
\sigma_t^{PK}(\vec{y}(0),\vec{x}(0))=:(\vec{y}(t),\vec{x}(t))\in 
\mathbb{R}^3\times \dot{\mathbb{R}}^3\ee
of the perturbed Kepler system we substitute 
$\vartheta(t)=\frac{1}{\sqrt{2}}(\eta(t)+i\xi(t))$ and 
$\zeta(t)=\frac{1}{\sqrt{2}}(i\eta(t)+\xi(t))$, where $(\eta(t), 
\xi(t))$ is given in (\ref{eta11}) for $l=1$   and in 
(\ref{eta1-1}) for $l=-1$,  into (\ref{KSy})-(\ref{KSx}). In consequence for $l=1$ we obtain
\begin{align}
x_1(t) & = \cos \left(\frac{\varphi_3'(t)}{2}\right)\cos \left(\frac{\Delta_1}{2}\right) - \sin \left(\frac{2\varphi_0(t)+\Delta_2}{2}\right)\cos \left(\frac{\varphi_3' (t)}{2}\right), \\
x_2 (t) & = \cos \left(\frac{\varphi_3'(t)}{2}\right)\cos \left(\frac{2\varphi_0 (t) + \Delta_2}{2}\right) + \sin \left(\frac{\varphi_3'(t)}{2}\right)\cos \left(\frac{\Delta_1}{2}\right),\\
x_3 (t) & = -\sin \left(\frac{\Delta_1}{2}\right) \cos \left(\frac{2\varphi_0(t) + \Delta_2}{2}\right), \\
y_1 (t) &= \frac{\cos \left(\frac{2\varphi_0(t) +\Delta_2}{2}\right)\cos \left(\frac{\varphi_3'(t)}{2}\right) }{I_0 - \sin \left(\frac{2\varphi_0(t) + \Delta_2}{2}\right) \cos \left(\frac{\Delta_1}{2}\right)},\\
y_2 (t) & = \frac{\cos \left(\frac{2\varphi_0(t) + \Delta_2}{2}\right)\sin \left(\frac{\varphi_3'(t)}{2}\right) }{I_0 - \sin \left(\frac{2\varphi_0(t) + \Delta_2}{2}\right) \cos \left(\frac{\Delta_1}{2}\right)},\\
y_3 (t) & = \frac{\sin \left(\frac{2\varphi_0(t) + \Delta_2}{2}\right)\sin \left(\frac{\Delta_1}{2}\right) }{ \sin \left(\frac{2\varphi_0(t) + \Delta_2}{2}\right) \cos \left(\frac{\Delta_1}{2}\right)- I_0} 
\end{align}
and for $l=-1$:
\begin{align}
x_1(t) & = \cos \left(\frac{\varphi_3'(t)}{2}\right)\cos \left(\frac{\Delta_1}{2}\right) + \sin \left(\frac{2\varphi_0(t) +\Delta_2}{2}\right)\cos \left(\frac{\Delta_1}{2}\right), \\
x_2 (t) & = \sin \left(\frac{\Delta_1}{2}\right)\sin \left(\frac{2\varphi_0 (t) + \Delta_2}{2}\right) - \cos \left(\frac{\varphi_3'(t)}{2}\right)\sin \left(\frac{\Delta_1}{2}\right),\\
x_3 (t) & = I_3'(t) -\sin \left(\frac{\varphi_3'(t)}{2}\right) \cos \left(\frac{2\varphi_0(t) + \Delta_2}{2}\right), \\
y_1 (t) &= \frac{\cos \left(\frac{2\varphi_0(t) + \Delta_2}{2}\right)\cos \left(\frac{\Delta_1}{2}\right) }{I_0 - \sin \left(\frac{2\varphi_0(t) + \Delta_2}{2}\right) \cos \left(\frac{\varphi_3'(t)}{2}\right)},\\
y_2 (t) & = \frac{\sin \left(\frac{2\varphi_0(t) + \Delta_2}{2}\right)\sin \left(\frac{\Delta_1}{2}\right) }{ \sin \left(\frac{2\varphi_0(t) + \Delta_2}{2}\right) \cos \left(\frac{\varphi_3'(t)}{2}\right)- I_0},\\
y_3 (t) & = \frac{\sin \left(\frac{2\varphi_0(t) + \Delta_2}{2}\right)\sin \left(\frac{\varphi_3'(t)}{2}\right) }{ \sin \left(\frac{2\varphi_0(t) + \Delta_2}{2}\right) \cos \left(\frac{\varphi_3'(t)}{2}\right)- I_0}, 
\end{align}
where $\varphi_0(t)$ and $\varphi_3'(t)$ are given in (\ref{519}) and (\ref{522}), respectively.

Substituting (\ref{eta11}) and (\ref{eta1-1}) into (\ref{IJ}) we find that the solution of the Hamilton equations (\ref{217}) is given by 
\begin{eqnarray}
I_1(t)&=\sqrt{I_0^2-(I_3'(t))^2}\cos\left(\frac{\varphi_3'(t)\pm\Delta_1}{2}\right),\\
I_2(t)&=\sqrt{I_0^2-(I_3'(t))^2}\sin\left(\frac{\varphi_3'(t)\pm\Delta_1}{2}\right),\\
J_1(t)&=\sqrt{I_0^2-(I_3'(t))^2}\cos\left(\frac{\varphi_3'(t)\mp\Delta_1}{2}\right),\\
J_2(t)&=\pm\sqrt{I_0^2-(I_3'(t))^2}\sin\left(\frac{\varphi_3'(t)\mp\Delta_1}{2}\right),\\
I_3(t)&=I_3'(t)=\mp J_3(t),\qquad\qquad\qquad\qquad
\end{eqnarray}
where the upper sign in front of $\Delta_1$ and $J_3(t)$ is for $l=1$ and the lower one for $l=-1$.

At the end let us mention that for $G_0 =0$ all the above formulas describe the solutions for the Hamilton equations defined by Hamiltonian (\ref{HK1}) for the Kepler system, see formulas (\ref{solK1}), (\ref{solK2}) and (\ref{solK3}).

\end{document}